\documentclass[12pt]{article}
\usepackage{fullpage,epsfig,graphics,amsbsy,amssymb}
\usepackage{graphicx}
\usepackage{amsfonts}
\usepackage{amssymb,amsmath,amscd}

 \newtheorem{proposition}{Proposition}

\newtheorem{example}[proposition]{Example}

\newcommand{\hol}[1]{{\bar{#1}}}
\newcommand{\ahol}[1]{#1}
\newcommand{\bea}{\begin{eqnarray}}
\newcommand{\eea}{\end{eqnarray}}
\newcommand{\beq}{\begin{eqnarray}}
\newcommand{\eeq}{\end{eqnarray}}
\newcommand{\nn}{\nonumber}

\def\M{{\cal M}}
\def\N{{\cal N}}

\begin{document}
\setcounter{page}{0}
\thispagestyle{empty}
\begin{flushright} \small
UUITP-16/09  \\
 \end{flushright}
\smallskip
\begin{center} \LARGE
{\bf   On the AKSZ formulation of  \\
the Rozansky-Witten theory\\
and beyond}
 \\[12mm] \normalsize
{\large \bf Jian~Qiu and Maxim Zabzine} \\[8mm]
 {
 \it
   Department of Physics and Astronomy,
     Uppsala university,\\
     Box 516,
     SE-75120 Uppsala,
     Sweden\\}
\end{center}
\vspace{10mm}

\begin{abstract}
 \noindent

   Using the AKSZ  formalism, we construct the Batalin-Vilkovisky master action for the
    Rozansky-Witten model, which can be defined for any complex manifold
     with a closed $(2,0)$-form.  We also construct the holomorphic version of Rozansky-Witten theory defined over
 Calabi-Yau $3$-fold.

\end{abstract}


\eject
\normalsize


 \eject

\section{Introduction}
\label{intro}

 In  \cite{Rozansky:1996bq} L.~Rozansky and E.~Witten introduced a $3$-dimensional topological sigma model
  whose target space is (compact or asymptotically flat) hyperK\"ahler manifold $M$.  We refer to
   this model as the Rozansky-Witten (RW) model.   The Feynman diagram  expansion of the partition function for
    this model gives rise to the finite type invariant of $3$-manifolds (Rozansky-Witten invariants) similar to those appearing
      in perturbative Chern-Simons theory.
       The reader may consult \cite{sawon}
        for the review of the  RW invariants and  \cite{Kapustin:2008sc} for the recent development in the RW-theory.

    In this work we  discuss the Batalin-Vylkovisky (BV) formulation of the RW model. Using
 the Alexandrov-Kontsevich-Schwarz-Zaboronsky (AKSZ)  prescription \cite{Alexandrov:1995kv}  we construct
   master action for $3$-dimensional topological sigma model which upon gauge fixing coincides with RW
    model.  The AKSZ-BV framework is conceptually powerful and it allows to address certain issues independently from
     a particular gauge fixing.
    Moreover the whole exercise is interesting on its own since it provides the relatively exotic example of
  the AKSZ construction.  This paper was inspired by the remarks from \cite{kontsevich1}.

 The paper is organized as follows. In Section \ref{AKSZ} we remind the basic aspects of
   AKSZ formalism.   Section \ref{RW-AKSZ} contains the AKSZ construction of $3$-dimensional
    sigma model with hyperK\"ahler target. In Section \ref{gauge} we demonstrate that upon the specific
  gauge fixing the BV model corresponds to the original formulation of RW model
   from  \cite{Rozansky:1996bq}.   As a simple corollary of the AKSZ formalism
      in Section \ref{holomorphic} we present the holomorphic version of RW theory defined over a Calabi-Yau 3-fold.
 Section \ref{end} contains the summary.

\section{AKSZ formalism}
\label{AKSZ}

In this Section we review the AKSZ construction \cite{Alexandrov:1995kv} of the solutions of the classical master equation
 within BV formalism.  We closely follow the presentation given in \cite{Roytenberg:2006qz} and we use the language
  of graded manifolds which are sheaves of $\mathbb{Z}$-graded commutative algebras over a smooth manifold,
   for further details the reader may consult \cite{Voronov:2001qf}. We consider both the real and complex cases and treat
    them formally on equal footing. However in complex case the additional care is required
     (see  \cite{Alexandrov:1995kv} for further details).

The AKSZ solution of the classical master equation is defined starting from the following data:

\medskip
\noindent{\bf The source}: A graded manifold $\N$ endowed with a homological vector field $D$ and
a measure $\int\limits_\N\mu$ of degree $-n-1$ for some positive integer $n$ such that the measure is
 invariant under $D$.

\noindent{\bf The target}:  A graded symplectic manifold $(\M,\omega)$ with $\deg(\omega)=n$ and a
homological vector field $Q$ preserving $\omega$.  We require that $Q$ is Hamiltonian,
 i.e. there exists a function $\Theta$ of degree $n+1$  such that $Q=\{\Theta,-\}$.  Therefore $\Theta$ satisfies
 the following Maurer-Cartan equation
 $$\{ \Theta, \Theta \}=0~.$$

Introduce the space of maps between ${\cal N}$ and ${\cal M}$,
${\rm Maps}(\N,\M)$ which should be understood
   as the space of morphisms of sheaves, which is again a sheaf over $C^\infty(\Sigma, M)$, where $\Sigma$ and $M$ are the
  underlying ordinary smooth manifolds of $\N$ and $\M$ respectively.  The space ${\rm Maps}(\N,\M)$ is naturally
   equipped with the odd symplectic structure. Moreover $D$ and $Q$ can be interpreted as homological vector
    fields acting on ${\rm Maps}(\N,\M)$ and preserving this odd symplectic structure. The AKSZ solution $S_{BV}$
      is the Hamiltonian for homological vector field $D+Q$  on ${\rm Maps}(\N,\M)$ and thus it satisfies automatically
       the classical master equation.

 Let us provide some details for this elegant construction.
Pick a map $\Phi\in{\rm Maps}(\N,\M)$. We choose a set of coordinates $X^A=\{x^\mu;\psi^m\}$ on
 the target $\M$, where $\{x^\mu\}$ are the coordinates for an open $U\subset M$ and $\{\psi^m\}$
 are the coordinates in the formal directions.  We also choose the coordinates $\{\xi^\alpha; \theta^a\}$ on the source
 ${\cal N}$, where   $\{\xi^\alpha\}$ are the local coordinates on $\Sigma$ and $\{ \theta^a\}$ are the coordinates in the formal
  directions of $\N$.
 The superfield $\Phi$ is defined as an expansion over the formal coordinates
  of $\N$ for $\Phi_0^{-1}(U)$
\begin{equation}\label{superfield}
\Phi^A= \Phi_0^A(u) +\theta^a  \Phi_{a}^A(u) + \frac{1}{2}\theta^{a_2}   \theta^{a_1}\Phi_{a_1a_2}^A(u)  + \ldots ~.
\end{equation}
The symplectic form $\omega$ of degree
 $n$ on $\M$ can be written in the Darboux coordinates $\omega = dX^A \omega_{AB} dX^B$.
 Using this form we define the symplectic form of degree $-1$ on ${\rm Maps}(\N,\M)$  as
\begin{equation}
 \label{P_structure}
 \omega_{BV} = \int\limits_\N \mu ~~\delta \Phi^A ~\omega_{AB}~ \delta \Phi^B~.
\end{equation}
 Thus the space of maps ${\rm Maps}(\N,\M)$ is naturally equipped with the odd Poisson bracket $\{~,~\}$.
 Since the space ${\rm Maps}(\N,\M)$ is infinite dimensional we cannot define the BV Laplacian properly. We can only talk about the naive Laplacian adapted to the local field-antifield splitting.
 However on ${\rm Maps}(\N,\M)$ we can discuss the solutions of the classical master equation. The AKSZ action then reads
\beq\label{AKSZ-action}
S_{BV}[\Phi]= S_{kin}[\Phi]+ S_{int}[\Phi] = \int\limits_\N \mu ~\left ( \frac{1}{2} \Phi^A\omega_{AB} D\Phi^B +
(-1)^{n+1} \Phi^*(\Theta)\right )
\eeq
and it solves the classical master equation $\{S_{BV},S_{BV}\}=0$ with respect to the bracket
 defined by the symplectic structure (\ref{P_structure}).
  In writing the solution we did not have to use the Darboux coordinate on $\M$. Assuming that
   $\omega$  admits Liouville form $\Xi$ the first term in (\ref{AKSZ-action}) can be written
    \beq\label{Liouvilleform}
   S_{kin} [\Phi] =  \int\limits_\N \mu ~ \Xi_A(\Phi) D\Phi^A~.
  \eeq
   Since the measure $\mu$ is invariant under $D$, $S_{kin}$ depends\footnote{Here we consider the case when $\partial \Sigma = \emptyset$. For the case with a boundary the reader may consult \cite{Cattaneo:2001ys}.}
    only on $\omega$, not a concrete choice of $\Xi$.
  The action (\ref{AKSZ-action}) is invariant under all orientation preserving diffeomorphisms of $\Sigma$
    and thus defines a topological field theory.  The solutions of the classical field equations of (\ref{AKSZ-action}) are graded differentiable maps $(\N, D) \rightarrow (\M, Q)$, i.e. maps which commute with the homological vector fields.

The homological vector field $Q$ on $\M$ defines a complex on $C^\infty(\M)$ whose
cohomology we denote  $H_Q(\M)$. Take $f\in C^\infty(\M)$ and expand $\Phi^*f$
in the formal variables on $\N$
$$\Phi^*f= O^{(0)}(f) +  \theta^a O^{(1)}_a(f) + \frac{1}{2} \theta^{a_2} \theta^{a_1}O^{(2)}_{a_1 a_2}(f)  + \ldots ~.$$
  We denote by $\delta_{BV}$  the
Hamiltonian vector field for $S_{BV}$, which is homological as a
consequence of classical
   master equation.
 The action of $\delta_{BV}$ on $\Phi^*f$ is given by the following expression
$$
\delta_{BV}(\Phi^*f)=\{S_{BV} ,\Phi^*f\} = D\Phi^*f + \Phi^* Qf~.
$$
Thus if $Qf=0$ and $\mu_k$ is a $D$-invariant linear functional on the functions of $\N$ (e.g.,
a representative of an homology class of $\Sigma$), then $\mu_k(O^{(k)}(f))$ is
$\delta_{BV}$-closed and can serve as a classical observable. Therefore $H_Q(\M)$ naturally defines
a set of classical observables in the theory. The classical action (\ref{AKSZ-action}) can be deformed to the first order by
$$ \int\limits_{\N}\mu~ O^{(n+1)}(f)$$
 with $f \in H_Q(\M)$.

The standard choice for the source is odd tangent bundle
$\N=T[1]\Sigma$, for any smooth manifold $\Sigma$ of dimension
$n+1$, with $D=d$ the de Rham differential over $\Sigma$ and the
canonical coordinate measure.

\begin{example} {\rm  ({\bf Chern-Simons theory})
 The Chern-Simons model is easily constructed within AKSZ framework \cite{Alexandrov:1995kv}.
  The source $\N = T[1]\Sigma_3$
  with de Rham differential and canonical integration. The target is $\M = \mathbf{g}[1]$ where $\mathbf{g}$ is a metric Lie algebra.}
 \end{example}

\begin{example} {\rm  ({\bf Poisson sigma model})
The AKSZ approach was applied to $2$-dimensional Poisson sigma model in \cite{Cattaneo:2001ys}. In this case
 the source $\N$ is  odd tangent $T[1]\Sigma_2$ for $2$-dimensional manifold $\Sigma_2$ equipped with de Rham differential
  and canonical integration measure. The target $\M$ is odd cotangent bundle $T^*[1]M$ with homological vector field associated
   to Poisson structure.}
 \end{example}

 \begin{example} {\rm ({\bf Courant sigma model})
 The AKSZ construction can be applied to $3$-dimensional Courant sigma model
  which associates for Courant algebroid a $3$-dimensional topological field theory,
   \cite{Ikeda:2002wh}, \cite{Roytenberg:2006qz}. The simplest example of this model corresponds to the following
    choice of $\N = T[1]\Sigma_3$ and $\M = T^*[2] T^*[1]M$.}
 \end{example}

 There exists other choices for the source supermanifold. For example, if $\Sigma$ is complex manifold with the holomorphic volume form then $\N =T^{(1,0)}\Sigma$ is equipped with integration and it is invariant under the homological vector field
  $D = \bar{\partial}$, the Dolbeault differential.

\begin{example} {\rm  ({\bf holomorphic Chern-Simons theory})
The source is $\N = T^{(1,0)}[1]\Sigma_6$ where $\Sigma_6$ is
Calabi-Yau $3$-fold. The measure has degree $-3$ and thus the
target $\M= \mathbf{g}[1]$ with
 $\mathbf{g}$ being a metric Lie algebra will work. It is simple exercise to check that the resulting theory would be the BV action
  for the holomorphic Chern-Simons theory \cite{Witten:1992fb}.}
 \end{example}

 The AKSZ prescription is algebraic in its nature and thus it  can be generalized  even further, see for
  example \cite{Bonechi:2009kx}.

 \section{Rozansky-Witten model from AKSZ}
 \label{RW-AKSZ}

 We are interested in the construction of $3$-dimensional topological sigma model.  If we choose   $T[1]\Sigma_3$ as a source
  manifold then the target $\M$ should be graded symplectic manifold of degree $2$. As in \cite{Roytenberg:2006qz} we consider   the even symplectic manifold $T^*[2] T^*[1] M$  with the symplectic structure
 \beq\label{symlplt2}
  \omega = \delta P_\mu \wedge \delta X^\mu + \delta v^\mu \wedge \delta q_\mu~,
  \eeq
  where the following allocation of degrees is assumed $\deg (X) =0$, $\deg (P) = 2$, $\deg (v) = \deg(q) =1$.
   If we assume that $M$ is a complex manifold then there exists a homological vector field $Q$ of degree $1$
   \beq\label{Qdefinieieiao}
    Q = P_{\bar{i}} \frac{\partial}{\partial q_{\bar{i}}}   + v^{\bar{i}} \frac{\partial }{\partial X^{\bar{i}}}~,
   \eeq
    where we use the complex coordinates $(i, \bar{i})$ on $M$.
 $Q$ preserves $\omega$ and the corresponding Hamiltonian is $\Theta = P_{\bar{i}} v^{\bar{i}}$.
 Picking up the local coordinates $X, P, v, q$ on $T^*[2] T^*[1] M$,  the space of maps
   $$ T[1] \Sigma_3 ~\longrightarrow~ T^*[2] T^*[1] M$$
   can be described by set of superfields $\boldsymbol{X}$, $\boldsymbol{P}$, $\boldsymbol{v}$, $\boldsymbol{q}$.
    For the components of $\boldsymbol{X}$ we use the following conventions
\beq\label{expansionfieldla44a}
  \boldsymbol{X}^\mu = X^\mu + \theta^a X_a^\mu + \frac{1}{2!}\theta^b \theta^a X^\mu_{ab} + \frac{1}{3!}\theta^c \theta^b \theta^a X^\mu_{abc}
 \eeq
  and the same conventions for other superfields.
 The space of maps is equipped with the odd symplectic structure
 \beq\label{BVoemmkkaka44}
    \omega_{BV} = \int d^3\theta d^3\xi~\left ( \delta \boldsymbol{P}_\mu \wedge
     \delta \boldsymbol{X}^\mu + \delta \boldsymbol{v}^\mu \wedge \delta
     \boldsymbol{q}_\mu  \right )~.
   \eeq
    Applying the AKSZ construction to this case we arrive on the following action
   \beq\label{masteraction1233}
S_{BV}=\int d^3 \theta d^3\xi  \left ( \boldsymbol{P}_\mu D\boldsymbol{X}^\mu +\boldsymbol{q}_\mu D\boldsymbol{v}^\mu
 +\boldsymbol{P}_{\bar{i}} \boldsymbol{v}^{\bar{i}} \right)~,
\eeq
  which automatically satisfies the classical master equation. This model is an example of family of $3$-dimensional
   sigma models associated to the generalized complex manifold (for further details see \cite{future}).
  However here we are interested in an exotic modification of this model.

  Assume that $M$ is complex manifold with a closed $(2,0)$-form $\Omega$.  In this case the symplectic structure
   (\ref{symlplt2})  on $T^*[2]T^*[1]M$ can be modified as follows
 \beq\label{22-02sisisi}
 \omega = \delta P_\mu \wedge \delta X^\mu + \delta v^\mu \wedge \delta q_\mu + \Omega_{ij} \delta X^i \wedge \delta X^j~,
 \eeq
 where we should require\footnote{Alternatively we can work only within $\mathbb{Z}_2$-grading.} that $\deg(\Omega)=2$.
  In other words we can think about introducing the formal parameter of degree $2$ and putting it in front of $\Omega$. Strictly
   speaking $T^*[2]T^*[1]M$ with (\ref{22-02sisisi}) is not a graded symplectic manifold since we have an auxiliary parameter
    with non-zero degree. However the AKSZ construction still goes through.  The homological vector field $Q$ in (\ref{Qdefinieieiao})
     preserves the new symplectic structure (\ref{22-02sisisi}) and the corresponding
    Hamiltonian is  $P_{\bar{i}} v^{\bar{i}}$. On the space of maps  $ T[1] \Sigma_3 ~\longrightarrow~ T^*[2] T^*[1] M$ the following
     odd symplectic structure is defined
   \beq\label{BVoemmkkaka}
    \omega_{BV} = \int d^3\theta d^3\xi~\left ( \delta \boldsymbol{P}_\mu \wedge
     \delta \boldsymbol{X}^\mu + \delta \boldsymbol{v}^\mu \wedge \delta
     \boldsymbol{q}_\mu + \Omega_{ij} \delta \boldsymbol{X}^i \wedge \delta \boldsymbol{X}^j \right )~.
   \eeq
   Writing locally $2\Omega_{ij} =  \partial_i \xi_j - \partial_j \xi_i$ with $\xi$ being $(1,0)$ holomorphic form
   we can apply the AKSZ construction and arrive at the following BV action
   \beq\label{masteraction12}
S_{BV}=\int d^3 \theta d^3\xi  \left ( \boldsymbol{P}_\mu D\boldsymbol{X}^\mu +\boldsymbol{q}_\mu D\boldsymbol{v}^\mu
 + \xi_i (\boldsymbol{X})  D \boldsymbol{X}^i +\boldsymbol{P}_{\bar{i}} \boldsymbol{v}^{\bar{i}} \right)~.
\eeq
 Despite its appearance the action (\ref{masteraction12})  depends only on $\Omega$ (not $\xi$) in the case $\partial \Sigma_3 =\emptyset$.
 The direct calculation shows
\bea
\{S_{BV},S_{BV} \}=\int d^3\theta d^3 \xi~ \left(D(\boldsymbol{P}_\mu D\boldsymbol{X}^\mu +
 \boldsymbol{q}_\mu D\boldsymbol{v}^\mu +\boldsymbol{P}_{\bar i}\boldsymbol{v}^{\bar i})+\Omega_{ij}
  D\boldsymbol{X}^i D\boldsymbol{X}^j\right)~.\nn
\eea
 The first term is immediately seen to be a surface term. To see
that the second term is also a surface term, we have to perform
the $\theta$ integrals
 \bea \int d^3\theta d^3\xi ~\Omega_{ij}
D\boldsymbol{X}^i D\boldsymbol{X}^j=\int d^3\xi
~\epsilon^{abc}\partial_c(\Omega_{ij}(\partial_aX^{i})X_b^{j})
\eea
 and use $d\Omega=0$.

 The model defined by (\ref{BVoemmkkaka44}) and (\ref{masteraction1233}) is formally related
   to the model defined by (\ref{BVoemmkkaka}) and (\ref{masteraction12}) through the formal shift $\boldsymbol{P}_i \rightarrow \boldsymbol{P}_i + \xi_i$ with $\deg (\xi)=2$ and $\xi$ being a holomorphic $(1,0)$ form such that $\Omega = d\xi$.
    Indeed there exists a whole family of models where $\Omega$-term in (\ref{BVoemmkkaka})
     and $\xi$-term in (\ref{masteraction12}) enter with
     the different numerical coefficients, which still satisfy the classical master equation.

\section{Gauge fixing of Rozansky-Witten model}
\label{gauge}

In the previous section we constructed the classical BV action for $3$-dimensional topological sigma model
 with the target being a complex manifold $M$ admitting the closed $(2,0)$ form $\Omega$.   In this section we
  discuss the gauge fixing of this model. In particular we show that when $M$ is hyperK\"ahler the gauged fixed
   version of (\ref{masteraction12}) is exactly the Rozansky-Witten model \cite{Rozansky:1996bq}.

 Let $M$ be hyperK\"ahler manifold with metric $g$ and holomorphic symplectic form $\Omega$ which
  is covariantly constant with respect to Levi-Civita connection. The gauge fixing in BV formalism consists
  of
    evaluating the BV action on the Lagrangian manifold. The main complication is related to the
  properties of $T^*[2]T^*[1]M$ (e.g.  see  \cite{Roytenberg:2002nu}).  $T^*[2]T^*[1]M$ is a vector bundle over vector
   bundle and thus $P$ transforms in non-tensorial fashion under the change of coordinates on $M$.   Moreover
    many components of superfields $\boldsymbol{X}$, $\boldsymbol{P}$, $\boldsymbol{q}$ and $\boldsymbol{v}$
      transform in rather complicated way.  The way out is the introduction of the connection and redefining some operations
       in the covariant way.  Let $\Gamma^{\nu}_{\mu\rho}$  be the Levi-Civita connection for  K\"ahler metric $g$.
         We redefine the coordinate of degree $2$ on
      $T^*[2]T^*[1]M$ and correspondingly the  superfiled
        $\boldsymbol{P}$ as follows
\bea
\mathbb{P}_{\mu}=\boldsymbol{P}_{\mu}+\Gamma^{\nu}_{\mu\rho}{\boldsymbol
q}_{\nu}{\boldsymbol v}^{\rho}~.
\eea%
 The master action (\ref{masteraction12}) becomes
\bea\label{dkdkdkk333}
S_{BV}= \int d^3 \theta d^3\xi   \left ( \mathbb{P}_\mu D\boldsymbol{X}^\mu +\boldsymbol{q}_\mu \nabla_D  \boldsymbol{v}^\mu
 + \xi_i (\boldsymbol{X})  D \boldsymbol{X}^i +\mathbb{P}_{\bar{i}} \boldsymbol{v}^{\bar{i}} \right)~,
\eea
  where
  \beq\label{eteu33399}
   \nabla_D \boldsymbol{v}^\mu = D\boldsymbol{v}^\mu + \Gamma^\mu_{\nu\rho} D \boldsymbol{X}^\nu \boldsymbol{v}^\rho~.
  \eeq
   In writing (\ref{dkdkdkk333}) we used the properties of the Levi-Civita connection for K\"ahler metric. We can also
    covariantize the $\theta$-derivatives
\beq\label{228993skk}
  \nabla_{\theta^a}:=\delta^{\mu}_{\rho} \partial_{\theta^a} + \Gamma^\mu_{\nu\rho}(\partial_{\theta^a}
  \boldsymbol{X}^\nu)
\eeq
and define the covariant components of the superfields. For example, we define
  \bea
   X_{ab}:=\nabla_{\theta^a}\partial_{\theta^b}\boldsymbol{X}|~,
 \eea
  where the vertical bar $|$ denotes "the $\theta = 0$ part". However we have to keep in mind that now $\nabla_{\theta^a}$
   and $\nabla_{\theta^b}$ do not anticommute.  The odd symplectic form (\ref{BVoemmkkaka}) can be rewritten in
    the covariantized superfields as follows
\bea\label{covarsympsl}
\omega_{BV} =\int d^3\theta d^3\xi  ~\left( \delta\left(\mathbb{P}_\mu\delta \boldsymbol{X}^\mu
 + \boldsymbol{q}_\mu \nabla_{\delta} \boldsymbol{v} ^\mu \right)+ \Omega_{ij} \delta \boldsymbol{X}^i \wedge \delta \boldsymbol{X}^j
  )\right)~,
\eea
 where
 \beq\label{codbdnal;a;als}
  \nabla_{\delta} \boldsymbol{v} ^\mu = \delta \boldsymbol{v}^\mu + \Gamma^\mu_{\nu\rho} \delta\boldsymbol{X}^\nu \boldsymbol{v}^\rho~.
 \eeq
 The $\theta$-integration of covariant scalar expression is defined as
\beq\label{2882sjjsjsjjs}
\int d^3\theta ... =  \frac{1}{6}\int  \epsilon_{abc}d\theta^ad\theta^bd\theta^c ...  =  \frac{1}{6}\int  \epsilon_{abc} \partial_{\theta^a}
\partial_{\theta^b} \partial_{\theta^c} ...  | = \frac{1}{6}\int  \epsilon_{abc} \nabla_{\theta^a}
\nabla_{\theta^b} \nabla_{\theta^c} ...   |
\eeq
 Now equipped with these tools we perform the gauge fixing by choosing the suitable Lagrangian submanifold for (\ref{covarsympsl})
  and evaluating the action (\ref{dkdkdkk333}) on it.

We now expand the symplectic form (\ref{covarsympsl})  in components
 and we shall ignore the $\boldsymbol{q}, \boldsymbol{v}$ sector
\bea\label{gopa23}
\omega_{BV}&=&\frac{1}{6}\int\epsilon^{abc}d\theta^a
d\theta^b\delta\left(
\partial_{\theta^c}\left(\mathbb{P}_\mu \delta
\boldsymbol{X}^\mu \right)+2\Omega_{\ahol{i}\ahol{j}} ( \partial_{\theta^c} \boldsymbol{X}^{i}) \delta \boldsymbol{X}^{j} + ... \right)\nn\\
&=&\frac{1}{6}\int\epsilon^{abc}\delta\big(({\cal
P}_{abc\mu}+2\Omega_{\ahol{i}\mu}X^{\ahol{i}}_{abc}+2\Omega_{\ahol{
i}\ahol{j}}X^{\ahol{i}}_cR(X_a,{}_{\mu},{}^{\ahol{j}},X_b)-3R(X_b,{}_{\mu},{\cal P}_a,X_c))\delta X^{\mu}\nn\\%
&&- (-3{\cal P}_{ab\mu}+6\Omega_{\mu\ahol{j}}X^{\ahol{
j}}_{ab})\nabla_{\delta}X^{\mu}_c+3{\cal P}_{c\mu} \nabla_{\delta}
X^{\mu}_{ab}+{\cal P}_\mu (\nabla_{\theta^a} \nabla_{\theta^b}
\nabla_{\theta^c}\delta \boldsymbol{X}^\mu)| + ... \big)~, \eea
 where $R$ is the curvature tensor is defined
 \bea
 [\nabla_{\mu},\nabla_{\nu}]v^{\alpha}:=R_{\mu\nu\ \beta}^{\ \
\,\alpha}v^{\beta}\nn\eea
and the components are
 $${\cal P}_\mu = \mathbb{P}_\mu|~,~~~~~~
 {\cal P}_{a\mu} = \nabla_{\theta^a} \mathbb{P}_\mu|~,~~~~~~
 {\cal P}_{ab\mu} = \nabla_{\theta^a} \nabla_{\theta^b} \mathbb{P}_\mu|~,~~~~~~
 {\cal P}_{abc\mu} = \nabla_{\theta^a} \nabla_{\theta^b} \nabla_{\theta^c} \mathbb{P}_\mu|~.$$
 From (\ref{gopa23}) we can pick the following Lagrangian
 submanifold
$${\cal P}_{\mu}=0~,~~~~~~
{\cal P}_{a\hol{i}}=0~,~~~~~~
{\cal P}_{a\ahol{i}}= 0~, ~~~~~~
{\cal P}_{ab\mu}=2\Omega_{\mu\ahol{j}}X^{\ahol{j}}_{ab}$$
$${\cal P}_{abc\mu}=-2\Omega_{\ahol{i}\mu}X^{\ahol{
i}}_{abc}
-2\Omega(X_a,{}_\ahol{j})R(X_b,{}_{\mu},{}^{\ahol{j}},X_c)$$
together with $\boldsymbol{q}=0$ (i.e., all components of
$\boldsymbol{q}$ are set to zero, which justifies why we ignored
the $\boldsymbol{q,v}$ sector from $\omega_{BV}$).
  The BV action (\ref{dkdkdkk333}) written in components is
\bea\label{wiwie88338}
S_{BV} &=&\frac{1}{6}\int d^3\xi ~\epsilon^{abc}\left(3{\cal
P}_{a\mu}\nabla_{[c}X_{b]}^{\mu}+3{\cal P}_{ab\mu}\partial_cX^{\mu}+6\Omega_{ij}
(X_c^{i}\nabla_bX_a^{j}+X_{bc}^{i}\partial_aX^{j}) \right . \nn \\
&& \left . +{\cal P}_{abc\hol{i}}v^{\hol{i}}+ {\cal P}_{ab\hol{i}}v_{c}^{\hol{i}} + {\cal P}_{a\hol{i}}v_{bc}^{\hol{i}}  + {\cal P}_{\hol{i}}v_{abc}^{\hol{i}}  + ...
\right)~,
\eea
 where dots stand for $\boldsymbol{q}_\mu \nabla_D \boldsymbol{v}^\mu$-term.
 Evaluating the action (\ref{wiwie88338}) on the above Lagrangian we recover the RW model
\bea\label{RWafetrgauge}
S_{RW}&=& \int d^3\xi ~\epsilon^{abc}\Big(
 \Omega(X_c,\nabla_bX_a)+\frac{1}{3} R_{\ahol{k}\hol{k}\ \ahol{j}}^{\ \ \
\ahol{
i}}X_b^{\ahol{k}}\Omega_{\ahol{l}\ahol{i}}X_a^{\ahol{l}}X^{\ahol{j}}_cv^{\hol{k}}\Big)\nn\eea%
or written in terms of differential forms on $\Sigma_3$
\bea S_{RW}&=&-6 \int  \left ( \Omega_{\ahol i\ahol j}X^{\ahol i}_{(1)}\wedge
d^{\nabla}X^{\ahol{j}}_{(1)} + \frac{1}{3} R_{\ahol{k}\hol{k}\ \ahol{j}}^{\ \ \
\ahol{ i}}X_{(1)}^{\ahol{k}}\wedge\Omega_{\ahol{l}\ahol{i}}
X_{(1)}^{\ahol{l}}\wedge X^{\ahol{j}}_{(1)}v^{\hol{k}} \right)~,
\eea
 where the only non-zero fields left in the model are odd $1$-form $X^i_{(1)} = X^i_a d\xi^a$,
  odd scalar $v^{\bar{i}}$ and even coordinate $X^\mu$.  The BRST transformations are obtained by restricting the BV
 transformation $\delta_{BV} \cdot=\{S_{BV},\cdot\}$ to the Lagrangian
submanifold
\bea
\{S_{BV},\boldsymbol{X}^{\hol{i}}\}=D\boldsymbol{X}^{\hol
i}+\boldsymbol{v}^{\hol{i}}&\Rightarrow&\delta
X^{\hol{i}}=v^{\hol{i}}~,\nn\\%
\{S_{BV},\boldsymbol{X}^{\ahol{i}}\}=D\boldsymbol{X}^{\hol
i}&\Rightarrow&\delta X^{\ahol{i}}=0\ \&\ \delta
X_{(1)}^{\ahol{i}}=- d X^{\ahol{i}}~,\nn\\%
\{S_{BV},\boldsymbol{v}^{\hol{i}}\}=D\boldsymbol{v}^{\hol
i}&\Rightarrow&\delta v^{\hol{i}}=0~.\nn\eea%
 The action (\ref{RWafetrgauge}) is invariant under these BRST transformation by construction \cite{Alexandrov:1995kv}.
For the sake of perturbation theory, we need a nondegenerate
kinetic term, this can be done through adding to the action (\ref{RWafetrgauge}) a BRST
exact term
\bea\label{exacttermsak}
S^{kin}_{RW}&=&-\delta\left(\int d^3\xi \sqrt{h} ~h^{ab}g_{i\bar{j}}X_a^{i}\partial_bX^{\bar{j}}\right)= \int d^3\xi \sqrt{h} \left (h^{ab}g_{i\bar{j}}\partial_aX^{i}\partial_bX^{\bar{j}}+h^{ab}g_{i\bar{j}}X_a^{i}\nabla_bv^{\bar{j}} \right ) \nn\\%
&=& \int g_{i\bar{j}}dX^{i}_{(0)}\wedge*dX^{\bar{j}}_{(0)}+g_{i\bar{j}}X_{(1)}^{i}\wedge
*d^{\nabla}v^{\bar{j}}~,\eea
 where we used the metric $h$ on $\Sigma_3$.  In (\ref{RWafetrgauge}) and
   (\ref{exacttermsak}) $d^\nabla$ is the covariant
  version of de  Rham differential, e.g.
  $$ d^\nabla v^\mu = d v^\mu + \Gamma^{\mu}_{\nu\rho}  dX^\nu v^\rho ~.$$
   The exact term (\ref{exacttermsak}) can be generated right away by a change of the Lagrangian submanifold
    by an appropriate gauge fixed fermion.

 In this section we analyzed the gauge fixing of the BV model introduced previously.
  In our setup we have used the K\"ahler metric $g$ and the fact that the holomorphic $(2,0)$-form
   $\Omega$ is covariantly constant with respect to the Levi-Civita connection.
The RW action above is by construction BRST invariant, but one can
nonetheless check this explicitly. For this one
  would  use some properties of the curvature tensor, some
of which are peculiar to a K\"ahler manifold and the covariant
constancy of $\Omega$, \bea \nabla_{[\rho}R_{\mu\nu]\ \alpha}^{\ \
\ \beta}=0;\ R_{ij\ \alpha}^{\ \ \beta}=0;\ R_{i\bar i\ j}^{\ \
\bar k}=R_{j\bar i\ i}^{\ \ \bar k};\ R_{\mu\nu\ [\ahol{j}}^{\ \
\;\ahol{l}}\Omega_{\ahol{i}]\ahol{l}}=0~.\nn\eea
 This setup is realized for a hyperK\"ahler manifold.
 One can give up the property of the metric being K\"ahler
   and the property of $\Omega$ being covariantly constant. In this case the analysis will be messier with a number of
    extra terms.
 Moreover, as far as BV formalism is concerned,  we do not  need to use anywhere that $\Omega$ is non-degenerate
  (i.e., it is a holomorphic symplectic structure).  However the present gauge fixing will lead to degenerate kinetic term for 
   the fermions. There may exists a different gauge giving rise to a well-defined perturbation theory and 
  thus leading to  a generalization of RW model to  any complex manifold
   with a closed $(2,0)$-form.

\section{Holomorphic Rozansky-Witten theory}
\label{holomorphic}

 From previous analysis we saw that the RW model corresponds to AKSZ construction with the source $T[1]\Sigma_3$
  and the target $T^*[2]T^*[1]M$ with the formal symplectic structure (\ref{22-02sisisi})  of degree $2$, where $M$ is
   complex manifold with  a closed $(2,0)$-form (e.g., $M$ can be hyperK\"ahler). The space of maps
   ${\rm Maps}(T[1]\Sigma_3, T^*[2]T^*[1]M)$  is equipped with the symplectic form (\ref{BVoemmkkaka}) of degree $-1$ since
    the source $T[1]\Sigma_3$ has a canonical measure of degree $-3$.  Thus the whole construction will work if we replace
     $T[1]\Sigma_3$ by another graded manifold equipped with homological vector field $D$ and  with invariant
      measure of degree $-3$.  For example, we can take $T^{0,1}[1]\Sigma_6$ where $\Sigma_6$ is a complex manifold with
       the holomorphic volume form. This choice of a source gives rise to the holomorphic version of RW model, very much
        in analogy with  holomorphic Chern-Simons theory introduced in \cite{Witten:1992fb}. Below we sketch the construction
         of the holomorphic RW theory. Our construction was inspired by the comments in \cite{kontsevich1}.

The source manifold is taken to be $T^{0,1}[1]\Sigma_6$ with
$\Sigma_6$ being a complex $6$-dimensional manifold with a holomorphic volume form $\Psi$ (i.e., $\Psi$ is
 nowhere vanishing closed $(3,0)$-form), which is written in complex coordinates $(z^a, \bar{z}^{\bar{a}})$ as
 $$\Psi = \rho (z)~ dz^1 \wedge dz^2 \wedge dz^3~,$$
  where $\rho(z)$ is a holomorphic density.  The integration on $T^{0,1}[1]\Sigma_6$ is defined as
  $$\int \rho~ d^3\bar{\theta} d^6\xi~ ... = \int \rho~ d^3 z ~d^3\bar{\theta} d^3 \bar{z}~... $$
   and it is of degree $-3$.
 On $T^{0,1}[1]\Sigma_6$ the homological vector field $D= \bar{\theta}^{\bar{a}} \partial_{\bar{a}}$ corresponds to
  the Dolbeault differential $\bar{\partial}$.  If $\partial \Sigma_6 =\emptyset$ the above measure is invariant under $D$.
 Thus most of the construction of the
holomorphic RW theory can be carried over from section \ref{RW-AKSZ}
by replacing the de Rham differential with $\bar\partial$. Thus on ${\rm Maps}(T^{(0,1)}[1]\Sigma_6, T^*[2]T^*[1]M)$  the
symplectic form is
   \beq
    \omega_{BV} = \int \rho ~d^3\bar\theta d^6\xi~\left ( \delta \boldsymbol{X}^\mu \wedge
     \delta \boldsymbol{P}_\mu + \delta \boldsymbol{v}^\mu \wedge \delta
     \boldsymbol{q}_\mu + \Omega_{ij} \delta \boldsymbol{X}^i \wedge \delta \boldsymbol{X}^j \right )
 \eeq
 and the master action is
 \beq
S_{BV}=\int \rho~ d^3 \bar\theta d^6\xi  \left (
\boldsymbol{P}_\mu  D\boldsymbol{X}^\mu +\boldsymbol{q}_\mu
D\boldsymbol{v}^\mu
 +  \xi_i D \boldsymbol{X}^i +\boldsymbol{P}_{\bar{i}} \boldsymbol{v}^{\bar{i}} \right)~,
\eeq%
which satisfies the classical master equation by construction.

 The gauge fixing of this model can be done in complete analogy with the real case described in section \ref{gauge}.
Skipping the details the gauge fixed action can be  written in terms of differential forms on $\Sigma_6$
\bea S_{hRW}&=&-6 \int \Psi \wedge  \left ( \Omega_{\ahol i\ahol j}X^{\ahol i}_{(0,1)}\wedge
\bar{\partial}^{\nabla}X^{\ahol{j}}_{(0,1)} + \frac{1}{3} R_{\ahol{k}\hol{k}\ \ahol{j}}^{\ \ \
\ahol{ i}}X_{(0,1)}^{\ahol{k}}\wedge\Omega_{\ahol{l}\ahol{i}}
X_{(0,1)}^{\ahol{l}}\wedge X^{\ahol{j}}_{(0,1)}v^{\hol{k}} \right)~,
\eea
 where the only non-zero fields left in the model are odd $(0,1)$-form $X^i_{(0,1)} = X^i_{\bar{a}} dz^{\bar{a}}$,
  odd scalar $v^{\bar{i}}$ and even coordinate $X^\mu$.  The BRST transformations are obtained by restricting the BV
 transformation $\delta_{BV} \cdot=\{S_{BV},\cdot\}$ to the Lagrangian
submanifold
$$ \delta X^{\hol{i}}=v^{\hol{i}}~,~~~~~~
 \delta X^{\ahol{i}}=0~,~~~~~~
\delta X_{(0,1)}^{\ahol{i}}= - \bar{\partial} X^{\ahol{i}}~, ~~~~~~
 \delta v^{\hol{i}}=0~.$$
 In order to have a well-defined perturbation theory we add to $S_{hRW}$ the BSRT-exact
 kinetic term
\bea
S_{hRW}^{kin} &=&-\delta\left(\int d^6\xi \sqrt{h} h^{\bar ab}g_{i\bar{j}}X_{\bar
a}^{i}\partial_bX^{\bar{j}}\right)=\int d^6\xi \sqrt{h} \left ( h^{\bar
ab}g_{i\bar{j}}\bar\partial_{\bar a}X^{i}\partial_bX^{\bar{j}}+h^{\bar ab}g_{i\bar{j}}X_{\bar
a}^{i}\nabla_bv^{\bar{j}} \right )~,
\nn\eea
 where $h$ is a Hermitian metric on $\Sigma_6$.  Indeed in order to have a well-defined kinetic term we have to
  require that $h$ is a K\"ahler metric\footnote{For the K\"ahler manifolds we have the relation for the different  Laplacians
    $\Delta_d= 2 \Delta_{\partial} = 2 \Delta_{\bar{\partial}}$ and this would allow us to have a well-defined propogator for
     the present kinetic term.} and thus $\Sigma_6$ is Calabi-Yau $3$-fold.

 In this section we have constructed the holomorphic RW model which is $6$-dimensional topological sigma model defined
  over Calabi-Yau $3$-fold with the hyperK\"ahler target. The perturbation theory for this model should give rise to holomorphic
   invariant of $3$-dimensional Calabi-Yau manifolds with the holomorphic volume form.

\section{Summary}
\label{end}

In this short note we analyzed the RW model and their generalizations within BV formalism.
 We used the elegant AKSZ-construction  which allows to construct the solution of
  classical master equation from simple geometrical data. The AKSZ treatment of RW model is a bit exotic
   example since we work with graded symplectic manifold with  the additional parameter ("coupling constant")
    with non-zero grading.

   AKSZ-BV framework is very powerful both conceptually and technically and many issues can be systematically
    addressed within this framework, such as boundary conditions for RW model, the coupling of RW model with
     Chern-Simons theory etc.

\bigskip\bigskip

\noindent{\bf\Large Acknowledgement}:
\bigskip

\noindent
 The research of M.Z. was
supported  by VR-grant 621-2008-4273.

\end{document}